# A new framework for global data regulation

Ellie Graeden[1*], David Rosado[1], Tess Stevens[1], Mallory Knodel[2], Rachele Hendricks-Sturrup[3], Andrew Reiskind[4], Ashley Bennett[5], John Leitner[6], Paul Lekas[7], Michelle DeMooy[8]

*Corresponding author

Affiliations
[1] Georgetown University, Center for Global Health Science and Security, Washington DC, USA
[2] Center for Democracy and Technology, Washington DC, USA
[3] Duke-Robert J. Margolis Center for Health Policy, Washington DC, USA
[4] Mastercard, New York City, New York, USA
[5] Independent Researcher, Seattle, Washington, USA
[6] SentiLink Corp., Washington DC, USA
[7] Software and Information Industry Association, Washington DC, USA
[8] Georgetown University McCourt School of Public Policy, Washington DC, USA

**Abstract**

Under the current regulatory framework for data protections, the protection of human rights writ large and the corresponding outcomes are regulated largely independently from the data and tools that both threaten those rights and are needed to protect them. This separation between tools and the outcomes they generate risks overregulation of the data and tools themselves when not linked to sensitive use cases. In parallel, separation risks under-regulation if the data can be collected and processed under a less-restrictive framework, but used to drive an outcome that requires additional sensitivity and restrictions. A new approach is needed to support differential protections based on the genuinely high-risk use cases within each sector. Here, we propose a regulatory framework designed to apply not to specific data or tools themselves, but to the outcomes and rights that are linked to the use of these data and tools in context. This framework is designed to recognize, address, and protect a broad range of human rights, including privacy, and suggests a more flexible approach to policy making that is aligned with current engineering tools and practices. We test this framework in the context of open banking and describe how current privacy-enhancing technologies and other engineering strategies can be applied in this context and that of contract tracing applications. This approach for data protection regulations more effectively builds on existing engineering tools and protects the wide range of human rights defined by legislation and constitutions around the globe.

**Data protection regulations**

Data are the abstract representation of the world and can now be used to describe nearly every aspect of our physician and digital experience. Smartwatches capture movement patterns and track other nearby watches (and the people who wear them) (1). Cars collect data on function and speed, alerting the driver when tire pressure is low and the insurance company when driving is erratic or dangerous (2). Credit card data collected by banking and finance apps and platforms can be used to provide access to financial accounts while individuals are home, traveling, marketing products and performing transactions within and across platforms (3). As data collection and use have expanded, data protection has become a topic at the forefront of discussion across much of the world (4). With a rapidly expanding number of artificial intelligence applications demonstrating the power of data processing and interpretation at scale, and as we are faced with increasing complexity and speed in technological development, there is an ever-growing and immediate need for robust, sophisticated governance to keep pace (5, 6).

For most of the world, data protections embedded at all levels of policy (i.e., local, state, provincial, national, institutional) have focused on the individual right to privacy, a long-standing legal foundation for data governance. While the focus on privacy regulation has intensified in the last few years, the legal basis for these policies have been in development for nearly 150 years.[1] The French Supreme Court recognized the right to protection of a private life in 1868 (7). The Right to Privacy, written by Samuel Warren and Louis Brandeis in 1890, was the first publication to argue for individual privacy legislation in the United States (US) (8, 9).[2] The modern Supreme Court found, 65 years later, in Griswold, that various amendments in the Constitution created a "zone of privacy," an implied right to privacy for Americans (10). Nearly 100 years later, in 1980, the Organisation for Economic Co-operation and Development (OECD) published Guidelines on the Protection of Privacy and Transborder Flows of Personal Data, the first major international protections focused specifically on privacy in data (11).[3] The OECD guidance was rapidly followed by the Convention for the Protection of Individuals with Regard to Automatic Processing of Personal Data ("Convention 108"), the 1981 treaty that established European policy on privacy, trade, and communications (12, 13). Convention 108 was recently modernized and aligned to the European Union's (EU) General Data Protection Regulation (GDPR), introduced in 2018 (14). GDPR was one of the first modern-era laws to protect individual privacy as a human right under a comprehensive omnibus law with enforcement for data protections. However, these protections lack clarity and do not adequately address the complexity of the data and technological landscape now in play, including the numerous and consequential ways in which data can be acquired and used (e.g., law enforcement, employment, insurance, etc.), leading to critical gaps in the regulatory framework.

---

[1] For the purposes of this paper, we use the terms laws, regulations, and policies interchangeably to refer to government action initiated and implemented by policies written and enforced by government actors.
[2] Even this early defense of privacy was triggered by technological developments. The advent of personal cameras allowed reporters to take and publish compromising photos, which embarrassed and set fear in the hearts of literati who preferred that their personal lives remain hidden from view.
[3] The Organisation for Economic Co-operation and Development (OECD) is an intergovernmental organization founded in 1961 to stimulate economic progress and world trade.

**The current model for regulating data protections**
Given the rapid expansion in the total volume of data and artificial intelligence models, there is an increasing imperative to transition from an atomistic approach to data protections to a systemic approach that accounts for the value chain of how data are used. Most critically, though privacy and data protection policies are centered on data, data do not stand alone. Far beyond their raw form, data from a wide variety of sources are collected, processed, and stored by tools. These tools subsequently yield outcomes that affect individuals and populations interacting with and impacted by how those data are used. These outcomes are the measured impacts in the world that can then be evaluated with the goal of safeguarding one or more human rights. Each of these elements – the data, tools, outcomes, and rights – can be envisioned as tiers of a pyramid, linked by use cases that extend from data to rights, though each component is currently regulated largely independently (Figure 1). Specific examples of regulations related to each tier are shown in Figure 1a.

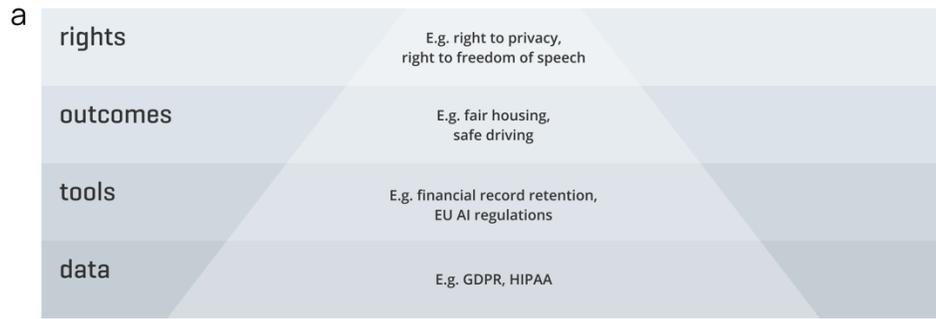
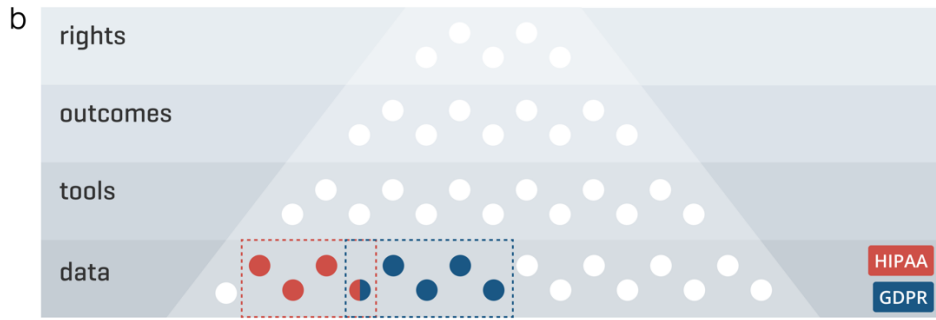
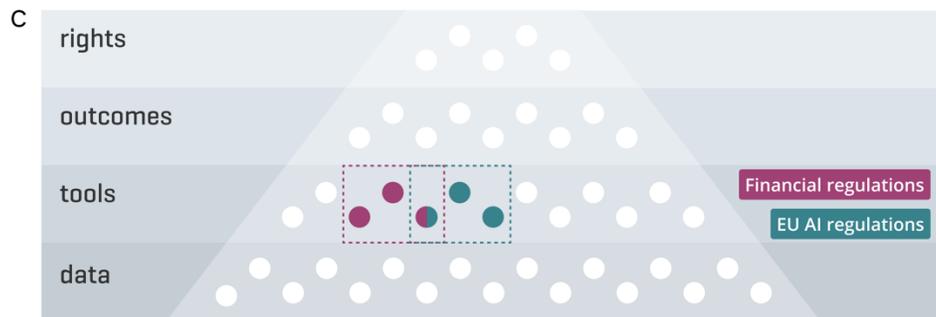
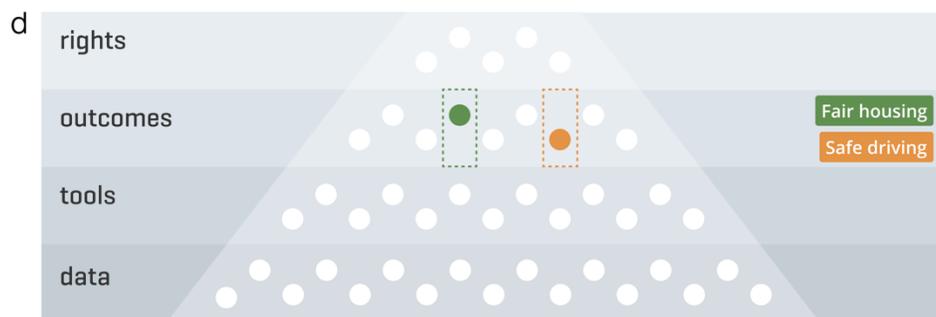
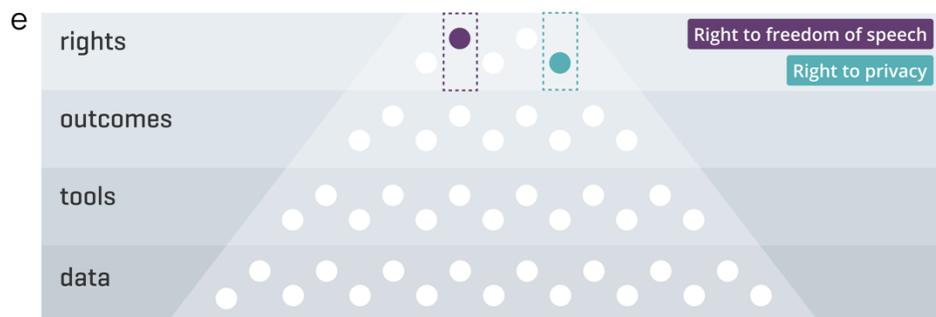

Figure 1. Current framework for regulating data protections. Each horizontal tier (shaded gray) represents a category of regulatory targets. For b-e, the white dots represent examples of what is regulated at each layer: specific datasets or categories of data in the data layer, specific tools or classes of tools in the tool layer, unique outcomes or impacts of how those data and tools are used in the outcomes layer, and specific rights in the rights layer. There are fewer dots at each layer as the pyramid narrows, representing the large number of, for example, data types as compared to rights. Specific examples are color-coded and described below.  a) Rights build on outcomes, which are generated by tools that use, store, and process data. Examples of policies are shown for each tier.  b) Current framework for regulating data. Red dots represent types of health data requiring protection under HIPAA; blue dots represent types of personal data that fall under GDPR; the red and blue dot represents a specific source or type of data that requires protection both under HIPAA and GDPR (e.g., consumer-generated health data, geolocation data, ride-sharing history data, etc.).  c) Current framework for regulating tools. Magenta dots represent tools that process or store financial data and fall under financial regulations; teal dots represent artificial intelligence models that are regulated under the new EU AI regulations; the magenta and teal dot represents an example of a model that processes financial data and also falls under the EU AI regulations. d) Current framework for regulating outcomes. The green dot represents a fair housing outcome; the orange dot represents outcomes regulated to driving safety. e) Current framework for regulation of rights. The purple dot represents the right to freedom of speech; the teal dot represents the right to privacy.

*Regulating data*

The current legal framework for data protections are largely based on the data themselves, the foundational tier in the pyramid. These protections typically fall into two categories: omnibus legislation and legislation specific to a single category of data. GDPR is a clear example of the former, as it targets protections for all personal data, a classification that is broad enough to capture nearly any data collected about people for any purpose (15). By contrast, the Health Insurance Portability and Accountability Act (HIPAA) regulates individually identifiable health information used or disclosed by defined 'covered entities,' which are limited to healthcare providers, insurance companies, and healthcare exchanges, while the Children's Online Privacy Protection Act (COPPA) regulates data collected, used, or disclosed online about a specific class of people: children under the age of 13, in this case (16).

Legal data protections tend to codify the data ecosystem of a given sector as it existed at the time of enactment: either by specific data types or by the industry or organization that collects and processes them (Figure 1b). For example, HIPAA was first developed in 1996 as part of the Social Security Act, and its protections were designed to allow individuals to maintain health insurance coverage between jobs. HIPAA covers health providers but does not extend to entities that, today, routinely access health data, such as mobile health apps and wearable providers leaving large gaps in data protections. Similarly, COPPA was enacted by Congress in 1998 as a way to make parents more aware of their children's online activities, a law that could not have anticipated the launch of the iPhone, and the corresponding seismic change it would bring to the digital world, just over a year later. Corresponding strategies for data protections based on changes in the digital space and technical characteristics are often then described in the regulation (e.g., specific strategies for de-identification) yielding a regulatory environment that can quickly become outdated as more effective privacy engineering solutions or privacy-enhancing technologies (PETs) are developed and new use cases for the data are implemented. Data regulations attempt to focus on the *how* of data protections; however, technological advancements are moving too quickly, data are expanding too rapidly, and data flows are too dynamic for the how to remain relevant for long. The disconnect in the rate of technological innovation versus that of developing policy to regulate those innovations has become an acute problem. Particularly as the amount of data expands, our knowledge of exactly which data were used by and for which tools decreases, and the tools used to analyze and model data generate even more data, amplifying the regulatory challenge exponentially. The current explosion of artificial intelligence models – from image generators to large language models and industry-specific models being developed in almost every field – have dramatically expanded not only the amount of data needed, but the amount generated. Under the current framework, each set of derivative results can require its own unique data protections, while the impacts and outcomes become harder to anticipate and evaluate.

*Regulating tools*
The tools that process, store, and transmit data are also currently regulated as standalone entities. The limitations placed on those systems are defined not by the specific function of the tool, but by its category as a tool to process data (Figure 1c). For example, the current proposed AI legislation from the EU would regulate AI models based on whether they are designed for and used in the context of high-risk applications (17, 18).[4,5] Similarly, financial regulations apply to any tools that store or process financial data, including policies on data retention (e.g., storage) (19) and security requirements for data transfer systems (20)[6].

Because current regulations are focused on the intended use of data and models, other outcomes influenced or impacted by the models that fall outside of the originally-identified target can be missed. For example, an algorithm designed to assess home values might not be categorized as high-risk, but could be considered high-risk if used in a different context, such as to identify newly gentrifying areas or to target policing to less affluent or distressed neighborhoods.[7] This type of "off-label" use would fall outside the current regulatory framework, despite the potential for harm. Conversely, this approach risks overregulation when tools are categorically defined as high-risk based on potential use versus specific use cases. For example, deploying a customer service chatbot in financial services is not the same level of risk as deploying machine learning to detect fraud, which could potentially increase discriminatory denial of services. Likewise, deploying a customer service chatbot to help insurance beneficiaries navigate their user portals online is substantially lower risk than using artificial intelligence to approve or deny prior authorization for potentially lifesaving health care services or treatments.

---

[4] The use of artificial intelligence will be regulated by the AI Act with key priorities to make AI safe, transparent and traceable. The Act requires such systems to be assessed based on the risk they pose to the users and require Generative AI to comply with transparency requirements.

[5] By contrast, the US National Institute of Standards and Technology's (NIST) Artificial Intelligence Risk Management Framework in January 2023 is a voluntary tool that is risk-based. However, coupled with a system engineering approach related to models and the data needed to train them, it provides flexibility for a broad range of organizations to achieve specified outcomes including privacy, security, safety, and fairness, which also encourages continual research and innovation to achieve more effective solutions for identified risks.

[6] EU-U.S. Data Privacy Framework. The framework fosters trans-Atlantic data flows and addresses security concerns.

[7] European Commission. "Regulatory framework proposal on artificial intelligence" High-risk AI are identified as technology used in critical infrastructure, education, safety components in manufacturing, employment, credit worthiness, law enforcement or democratic processes.

Conversely, focusing on the tools themselves means that regulations rarely account for the data used to train the model or stored by the system. Without taking the data types into account, regulations on tools can work at cross-purposes. For example, a new law proposed in Utah, requires that social media companies verify age to ensure children cannot open new accounts.[8] In practice, however, this regulation could significantly increase the amount of sensitive personal data accessed or collected by the companies in the process of verifying age (e.g., by requiring upload of an image of a government identification card). This collection could be lessened if privacy-enhancing technologies (PETs) suitable for identity use cases are deployed, but in many cases these types of engineering solutions are not addressed or supported as a risk mitigation tool because the regulatory approach is focused solely on the data.

*Regulating outcomes*
Most laws implemented by governments globally are focused on the outcome or impact of actions by individuals or institutions. Examples of regulations focused on outcomes include laws targeting housing discrimination, vehicle safety, child welfare, and many other areas (Figure 1d). Unlike the policies regulating data and tools, regulations and laws targeting outcomes tend to be narrower, focused on context-specific domains. For example, housing discrimination is illegal, whether implemented using a paper map and a red pen (i.e. red-lining) or using race-based credit worthiness algorithms (21).

In another, more specific example, the Equal Credit Opportunity Act states that individuals applying for credit can only be evaluated using factors related to their creditworthiness and prohibits any form of discrimination, such as on the basis of race, gender, color, religion, or age (22). This framing focuses on the data sources for establishing creditworthiness, not the outcome itself. However, even when specific identifying characteristics of the individual are not included in the model, other proxy variables can often be used that yield the same result. Even a few data elements about the individuals' digital footprint such as the type of device, operating system, time, email, and email domain are correlated with protected classes and are often used as measures of creditworthiness (23). If, instead, the Act were oriented around the outcome of biased assessments of creditworthiness, the law could be more effectively applied to the rapidly evolving technologies that are yielding discriminatory outcomes despite meeting the letter of the law in terms of the data used. Impact should be the critical measure of whether a law is broken, not the data and tools used.

---

[8] The new Utah laws—H.B. 311 and S.B. 152—require that social media companies verify the age of any Utah resident who makes a social media profile and confirm parental consent for any minor who wishes to make a profile.

*Regulating rights*

Protecting human rights is a primary goal of law and policy. However, when we implement protections for each right independently, we risk privileging one right over another unintentionally – decreasing protections for one as an unintended effect of increasing protections for another (Figure 1e). Even in the earliest cases of privacy in the courts, such as those litigated by Brandeis, there was already a gray area between the right to privacy and the right to free speech and freedom of the press. The line between what can or cannot be published about civilians has been deemed different from what can be published about public personas or those running for office: that which might otherwise be considered private information is considered of the public interest if the person is up for election.[9] It is these very conflicts that are the critical purview of legal scholarship and regulatory application. In a 2023 example, the Federal Trade Commission (FTC) sued GoodRx for violating the FTC Act and the Health Breach Notification Rule, for inappropriate sharing of health data as though it were consumer data, stating that the company had violated HIPAA by using health-derived data for targeted advertising (24). While the case was settled before it went to court, the challenge from the FTC highlights the regulatory ambiguity for health-adjacent data that could arguably be defined as either consumer or health, which dramatically changes the legal framework under which the data are regulated. The challenge facing policy makers is how to effectively protect the privacy of health data, fair use of personal data, and the right to consumer protections related to the products we buy. Without a clear regulatory framework that differentiates between the use of data in each context, these rights become conflated and their protection diluted.

Since early privacy legislation was developed, the way we collect and share information about the world around us and about each other has changed dramatically. The question now is whether or not our current regulatory framework, focused on the regulation of **data** by category and limitations on the **tools** that store and process those data, adequately protects the **outcomes** and **rights** that we want to protect. And, if they do not adequately address these outcomes and rights, how do we shift the framework to protect not just privacy, but the wide range of human rights described in global constitutions from the right to free speech and press to equal protection under the law, from fair markets to the right to meet our basic needs for housing, food, and health?

**A new regulatory paradigm**

Under the current regulatory framework for data protections, the protection of human rights writ large and the corresponding outcomes are regulated largely independently from the data and tools that both threaten those rights and are needed to protect them. This separation between tools and the outcomes they generate risks overregulation of the data and tools themselves when not linked to sensitive use cases. In parallel, separation risks under-regulation if, as in the GoodRx case, the data can be plausibly collected and processed under a less-restrictive framework, but used to drive an outcome that requires additional sensitivity and restrictions. A new approach is needed to support differential protections based on the genuinely high-risk use cases within each sector.

---

[9] The Ethics in Government Act of 1978 requires high-level federal officials to publicly disclose their personal financial interests. The public filing of this information is intended to prevent financial conflicts of interest.

Here, we propose a new framework in which data protection regulations are organized vertically to capture the entire value chain of data use, from the data and tools to their applied use cases, outcomes, and associated rights (Figure 2). Instead of each horizontal layer being regulated individually, this regulatory paradigm shifts the emphasis toward how data are used in specific contexts as a part of a process, and away from their regulation as a good or bad outcome in and of themselves.

By shifting to a vertically-aligned regulatory framework that honors both the context and direction in which data is collected, used, or shared, we gain immense flexibility to limit the use of data and tools in one context for one outcome while allowing the use of those same data and tools under a different regulatory framework to drive toward a different outcome, all while protecting a wide range of rights. This model (Figure 2b) would support, for example, the use of health data in the aggregate for public health response efforts while limiting their use to assess insurability. By establishing regulations based on use or outcome, the tools and the data that drive them can be used to benefit population-level health while still protecting the human right to health and healthcare in addition to the right to privacy (25). Conversely, a single right or multiple rights can be more effectively protected when linked to specific use cases for the data and models that drive a diversity of outcomes (see Figure 2c).

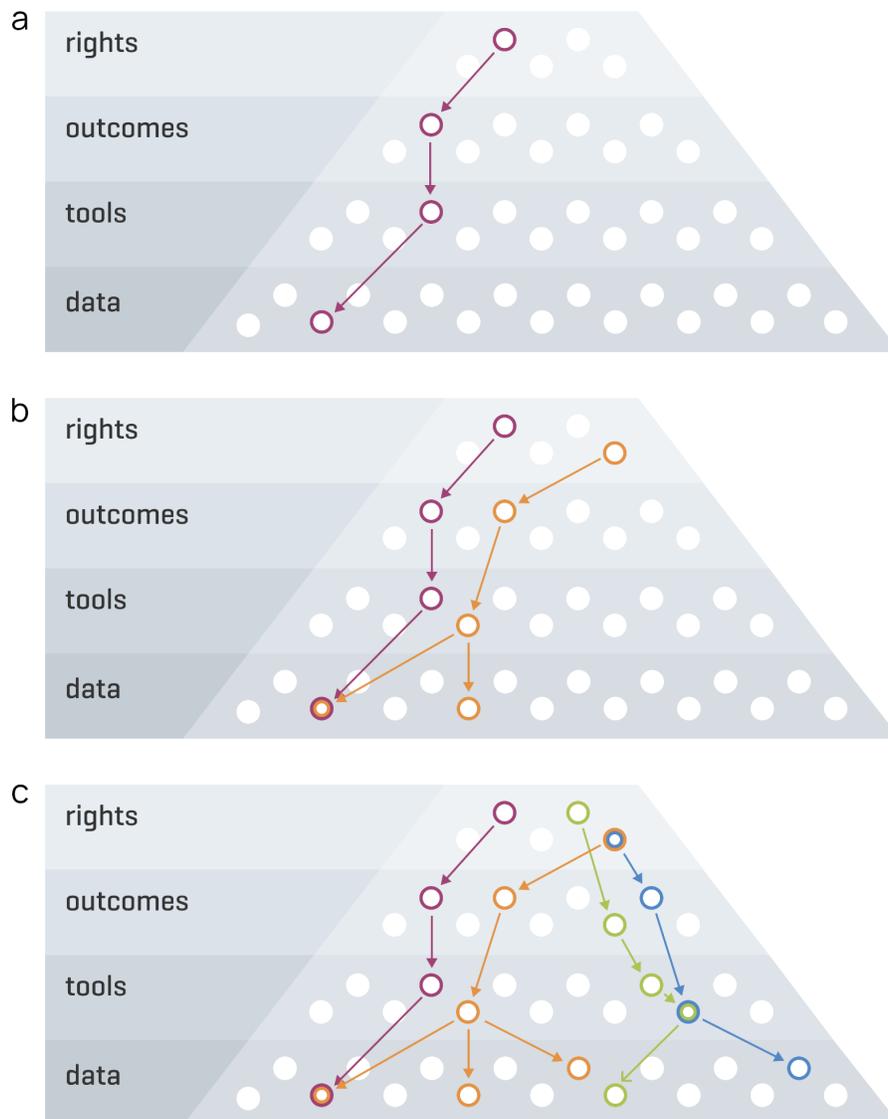

Figure 2. A vertically-aligned paradigm for regulating data protections. Each horizontal gray shaded tier represents a category of regulatory targets. Each colored line shows one example of a right, linked to one or more white dots which represent outcomes with data processed by one or more tools from one or more data sources. Each color represents a different regulatory framework based on a different combination of elements from each tier. a) A single right and associated outcome drives the regulation of tools and data that are used to achieve that outcome. b) Different outcomes fall under separate regulations, though the data tools used to achieve those outcomes may overlap. c) A representation of the new framework showing multiple linked regulatory targets, accounting for heterogeneity and multi-use of data and models while organizing regulation around the protection of diverse rights and corresponding outcomes.

**Applying the framework**

A vertical framework for data protection regulations requires a significant intellectual shift. However, the existing regulations and engineering methods needed to implement the new approach are readily available and already in place in many cases. Below, we describe one example for how this framework can be applied.

*Open banking*

Financial regulations are highly prescriptive, responsible for governing the movement of data globally for more than 70 million transactions daily between individuals and organizations (26). Open banking describes the provisions that regulate this process, ensuring the data–and the money–can be managed and transferred securely and privately between any set of financial service providers upon request by the customer (3). Currently, the data, the tools to process those data, and the outcomes of how the data are used are regulated independently.

The financial and personal **data** required for open banking can include account balances; types of institutions/entities involved in the transaction; transaction geolocations; personal information like names, addresses, and social security numbers; measures of creditworthiness; and more. These data are regulated under highly-prescriptive consumer privacy protections (27). Similarly, the **tools** that handle and transfer data currently need to satisfy uptime specifications, access control mechanisms, and encryption standards, including record retention, and are regulated by a large number of national and international policies, treaties, and agreements all of which define the requirements for open banking services.[10,11,12] The **outcomes** of open banking are regulated independently, from the ability to purchase goods directly from your bank account to the ability of consumers to dispute their credit scores and the responsibility to report financial fraud (22, 28, 29).[13] These outcomes impact and support a wide range of individual **rights**, from the right to own and exchange your purchases, to consumer protections, to the right to privacy. The protection of additional rights, including protection from corruption, fraud, and other illegal activities masked by money laundering such as human trafficking and the contraband of endangered species and drugs, require systems-wide analysis of large amounts of population-level data, which can conflict with personal privacy legislation in the financial domain.

If we shift the legal framework to a top-down vertical approach based on specific outcomes that meet specific rights, we can apply a more systems-based approach to risk mitigation and protection of both individual and societal rights. For example, the right to access personal funds to make purchases is currently treated as a standalone right. However, that right is predicated on a system that links the right to the outcome: the ability to make a point of sale purchase. Given a vertical framework, the **right** to access personal funds at the point of sale and the **outcome** of successfully completing a purchase can be linked to a **tool** that performs fraud detection. Fraud detection tools require access to anonymized **data** about typical transactions by similar customers to provide the statistical foundation for anomaly detection as well as personal information about the individual's spending and travel habits. This specific combination of information, while needed for rapid assessment of point-of-sale transactions, would not be relevant nor should it be accessible or used to assess mortgage creditworthiness.  In the case of mortgage creditworthiness, the individual **right** to fair access to housing, as supported by an **outcome** of bias-free assessment of creditworthiness is driven by models (**tools**) that, like fraud detection, require aggregate **data** about the general population and specific information about the individual applying for credit. While the data used are similar and require similar privacy protections in each case, a model of mortgage creditworthiness requires significant bias testing (in the United States, under the Federal Housing Act) while fraud detection does not. By structuring the data protection regulations vertically, the rights and associated outcomes in each case can be met, while preventing overregulation and burdensome requirements that are decoupled from the relevant outcome and associated rights.

---

[10] 15 U.S.C. 6801(b), 6805(b)(2) Part 314. GLBA's Safeguard Rule "sets forth standards for developing, implementing, and maintaining reasonable administrative, technical, and physical safeguards to protect the security, confidentiality, and integrity of customer information."

[11] Federal Trade Commission. Disputing Errors on your Credit Report. "As long as the information is correct, a credit bureau can report most negative information for seven years, and bankruptcy information for 10 years."

[12] The Revised Payment Services Directive (PSD2) is a European directive, administered by the European Commission to regulate payment services and providers throughout the European Union and the European Economic Area.

[13] The Currency and Foreign Transactions Reporting Act of 1970 -commonly referred to as the "Bank Secrecy Act" (BSA) - requires U.S. financial institutions to assist U.S. government agencies to detect and prevent money laundering.

**Engineering for a vertical regulatory framework**
The engineering tools needed to build rights-protecting data platforms are already available. These strategies, including PETs, can be used not only to manage risk associated with data storage and transfer, but to minimize the data collected in the first place (30). Notably, privacy engineering is a systems-wide approach with linked methods applied to the data and tools to generate specific outcomes associated with user requirements. Therefore, aligning the regulatory framework to vertically-oriented technical applications amplifies the value of the existing engineering tools and strengthens the implementation of both the regulations and engineering strategies.

Contact tracing for infectious disease outbreaks provides a useful example of outcomes-oriented privacy engineering (31). The success of global outbreak response depends largely on the ability to effectively and rapidly share the data needed to address both individual and public health. Contact tracing applications, mobile software designed to collect information about infection, were launched in many countries during the Covid-19 pandemic to integrate user location data with test results showing infection status. These data were used by individuals to assess their risk of infection and also shared by governments or other public health officials to assess risk across the population or in specific communities (32). For example, the Exposure Notifications System introduced jointly by Apple and Google is a privacy-preserving contact tracing application specifically engineered to alert users of potential exposures. The application enabled public health authorities to collect aggregated data to monitor the evolution of the pandemic, while upholding strong privacy principles (33). The architecture used in the Exposure Notifications System builds on cryptographic secure aggregation and differential privacy, and relies on proximity to another infected user, not location tracking, to minimize data collection. When coupled with encryption during data transfer and differential privacy to limit access, these methods significantly reduced the risk of data transfer in good part by limiting the amount of information that needed to be transmitted between systems. By linking methods applied to both the data and tools, the applications could more effectively and safely address outcomes required for both individual and public health response.

Given the current regulatory framework, these systems were subject to a long list of regulations, each based on the type of data and the way it was shared. HIPAA and related global health data regulations applied to data that were collected by applications managed or made available by healthcare providers or insurers; COPPA applied specifically to data collected about children; GDPR applied to data collected about European Union citizens; and both the Communications Act and Electronic Communications Privacy Act applied to data collected on telephones or other regulated electronic devices (34). If the proposed vertical regulatory framework were applied instead, we could regulate these data and tools based on the specific use case–individual health and privacy or population-level well-being. The new approach would build on the vertically-integrated engineering tools that collect and process the data, protect the right to individual privacy, and more effectively protect the right to health by supporting and ensuring access to population-scale data collected from individuals.

This vertical framework, applied to engineering data protections, is closely aligned with and would support the application of new outcomes- and risk-based guidance published by the National Institute of Standards and Technology and the latest artificial intelligence regulations proposed by the European Union. Both the guidance and the regulations are organized around the use case of the models and the sensitivity of the data used to train them. Starting with outcomes ranked by risk, each model and the underlying data it uses are then required to meet different standards, driving a more systems-wide approach to regulation. Expanding this approach beyond artificial intelligence models into the broader technology domain would significantly reduce the costs to implementation and innovation while more effectively protecting all human rights.

**Conclusion**
Current privacy regulations are focused narrowly on the data and associated tools. This regulatory approach risks prioritizing privacy over other equally critical human rights and, from a tactical perspective, the current legal structures have failed to keep up with the pace of technological development. By designing regulation for data and related tools as we do in most other regulatory contexts–governing outcomes–we protect our diverse human rights and the outcomes that are the expression of those rights. While this shift requires a significant change in how data regulations are structured, the engineering strategies needed to implement the new approach are already in place and, indeed, better aligned to a vertical, systems-based approach to regulation than the current regulations that treat each tier independently. By changing our focus from regulating data and tools to regulating the outcomes and uses of those data and tools, we can build a regulatory framework that is flexible, enduring, and effective.

**Acknowledgements**
This work was developed in collaboration with participants in the Georgetown Data Policy and Engineering Symposium held on February 9, 2023 at Georgetown University, including Ashley Bennett, Aneesh Chopra, Alissa Cooper, Marc Crandall, Ryan Donaghy, Ellie Graeden, Rachele Hendricks-Sturrup, Mallory Knodel, Naomi Lefkovitz, John Leitner, Paul Lekas, Kobbi Nissim, Pamela Peele, Andrew Reiskind, and Rob Sherman. The symposium was supported by the Georgetown research team at the Center for Global Health Science and Security of Hailey Robertson, David Rosado, Tess Stevens, and Ryan Zimmerman. The Symposium was funded by a gift from Meta.